\documentclass[conference]{IEEEtran}

\ifCLASSINFOpdf
\else
   \usepackage[dvips]{graphicx}
\fi
\usepackage{url}

\usepackage{graphicx}
\usepackage{booktabs,multirow,pifont,hyperref}

\begin{document}

\title{ProtoKD: Learning from Extremely Scarce Data for Parasite Ova Recognition}

\author{Shubham Trehan$^1$, Udhav Ramachandran$^2$, Ruth Scimeca$^2$, and Sathyanarayanan N. Aakur$^1$\\
$^1$ Department of Computer Science and Software Engineering, Auburn University, Auburn, AL, 36849\\
$^2$ Department of Veterinary Pathobiology, Oklahoma State University, Stillwater, OK, 74078}

\maketitle

\begin{abstract}
Developing reliable computational frameworks for early parasite detection, particularly at the ova (or egg) stage, is crucial for advancing healthcare and effectively managing potential public health crises. While deep learning has significantly assisted human workers in various tasks, its application in diagnostics has been constrained by the need for extensive datasets. The ability to learn from an extremely scarce training dataset, i.e., when fewer than 5 examples per class are present, is essential for scaling deep learning models in biomedical applications where large-scale data collection and annotation can be expensive or not possible (in case of novel or unknown infectious agents). In this study, we introduce ProtoKD, one of the first approaches to tackle the problem of multi-class parasitic ova recognition using extremely scarce data. Combining the principles of prototypical networks and self-distillation, we can learn robust representations from only one sample per class. Furthermore, we establish a new benchmark to drive research in this critical direction and validate that the proposed ProtoKD framework achieves state-of-the-art performance. Additionally, we evaluate the framework's generalizability to other downstream tasks by assessing its performance on a large-scale taxonomic profiling task based on metagenomes sequenced from real-world clinical data. 

\end{abstract}


\IEEEpeerreviewmaketitle

\section{Introduction}
Parasitic infections pose a significant threat to human and animal health, often leading to severe illness and even fatalities. These infections can be transmitted through various means, including contaminated food and water sources and common disease vectors such as mosquitoes. For instance, certain zoonotic diseases can be transmitted to humans through the consumption of infected livestock, such as cows and pigs. 
Many infections, particularly gastrointestinal, can cross the species barrier, and can further amplify the public health risks associated with them. Detecting these parasites early, especially at the ova stage, is crucial for preventing outbreaks of parasitic diseases.
Parasitic ova (eggs or cysts) have unique characteristics that allow them to be a distinguishing factor between different kinds of parasitic infections. 
The typical identification process requires the isolation of the egg from fecal samples collected from the infected host which are subsequently analyzed by a parasitologist under a microscope. 
Genus-level identification can enable the development of treatments to prevent severe infections and large-scale outbreaks. 

\begin{figure*}[t]
    \centering
    \includegraphics[width=0.8\textwidth]{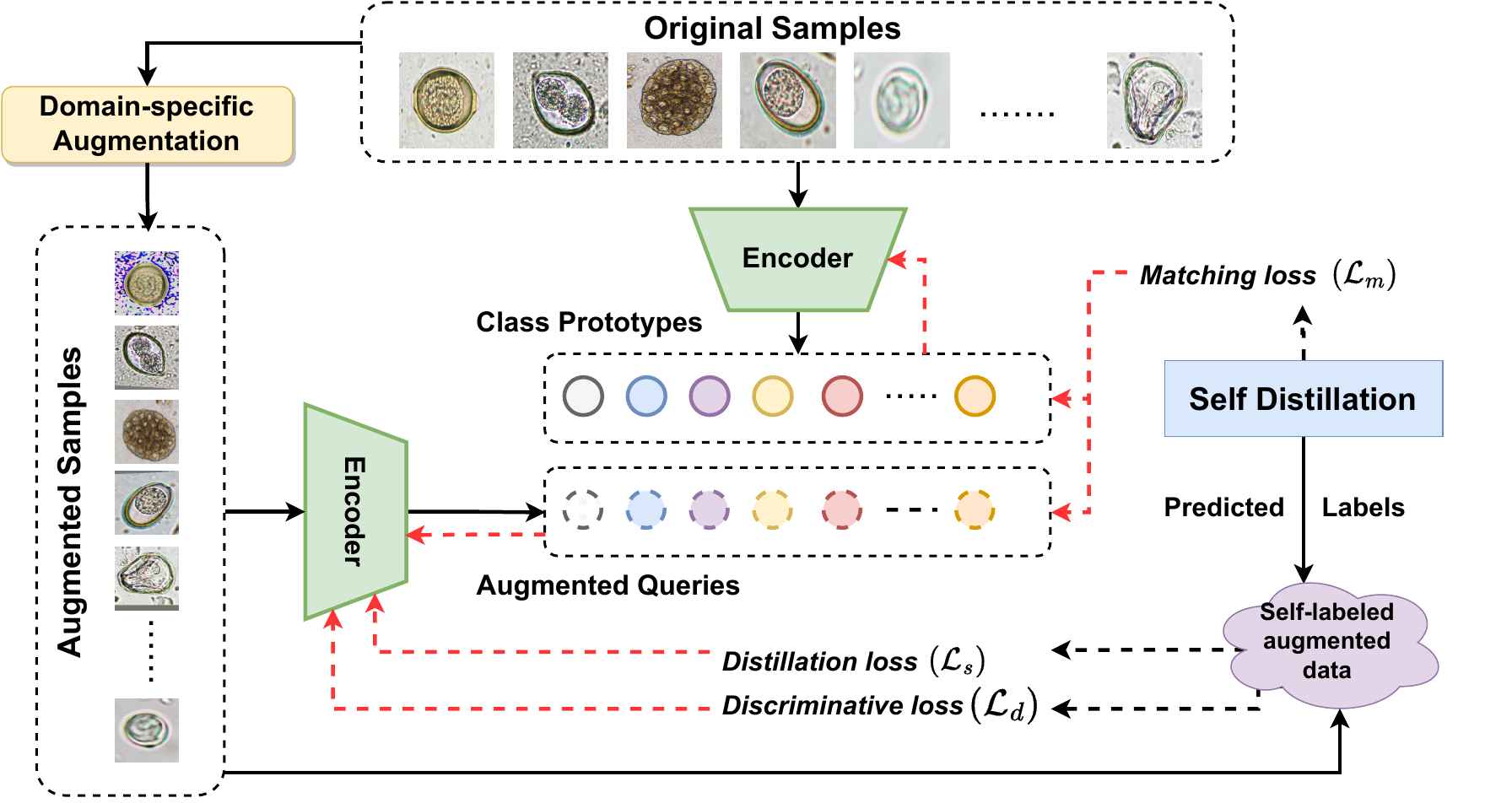}
    \caption{The \textbf{overall architecture} of the proposed ProtoKD is illustrated. Based on prototypical networks and self-distillation, a two-stage process captures the intra-class variations that can occur in biological data in a robust representation.}
    \label{fig:arch}
\end{figure*}

Deep learning has enabled the development of ``smart'' systems that have made immense progress in many fields. However, enormous amounts of historical data (hundreds if not thousands of samples per class) have been necessary to derive insights for downstream applications. Advances in disease diagnostics have been sparse due to the need to learn associations from highly limited, constrained data. Acquiring clinical data to train such deep learning models can be expensive, both in terms of the data acquisition cost and the time and effort of highly skilled human users to provide high-quality, large-scale annotations, which curbs the applications of deep learning models trained under a traditional supervised setting. Transfer learning~\cite{ghafoorian2017transfer} and few-shot learning~\cite{finn2017model,NIPS2016_90e13578,https://doi.org/10.48550/arxiv.1703.05175} have somewhat alleviated this dependency but still require significant amounts of quality data. Generative models such as diffusion models~\cite{ho2022cascaded} and generative adversarial networks (GANs)~\cite{goodfellow2020generative} show potential in generating synthetic samples for training data augmentation but have been prone to memorizing and replicating training data~\cite{corneanu2020computing}, and prone to hallucinating artifacts~\cite{cohen2018distribution}. Hence, their application in biomedical settings is inhibited due to privacy concerns~\cite{fredrikson2015model}, where patient identity and data integrity could be compromised. 

In this work, we present \textit{ProtoKD}, a framework designed to work with extremely scarce data, e.g., where less than $5$ training samples are present per class. Such settings are common in biomedical and diagnostic applications where labeled training data can be expensive. There is a need for rapid learning from a few samples for novel or unseen classes of interest. 
An illustration of the approach is shown in Figure~\ref{fig:arch}. Our approach was based on the idea that learning robust representations of limited training data and using \textit{domain-specific} augmentations to create an auxiliary dataset that can, together, capture intra-class variation.
Through a cyclical, two-phase process, we aim to align representations from original and augmented images to capture variations in the decision boundary across closely related classes. 
First, a matching loss is introduced to learn robust representations to distinguish \textit{between} classes using a prototypical network. 
Second, a self-distillation loss is introduced to help capture the intra-class variations in the data by presenting the network with heavily augmented data and training with pseudo-labels generated by the prototypical network. This step has a two-fold effect: (i) it adds a level of regularization that prevents the networks from overfitting, and (ii) it allows us to introduce other learning losses that help discriminate between fine-grained representations. 
By extending the idea of prototypical networks and self-distillation, we learned robust representations from extremely sparse data that was capable of capturing the intra-class variations through domain-specific data augmentation. 

The \textbf{contributions} of our work are three-fold: (i) we present one of the first works to address the problem of multi-class parasitic ova recognition from microscopic images, (ii) we develop a framework to learn from extremely scarce data (from a single example per class) in a multi-class classification setting, and finally (iii) we demonstrate its generalization to other biomedical tasks by evaluating on metagenome profiling.

\subsection{Related Work}
There have been very few \textbf{automatic parasitic ova detection} frameworks explored in literature. Supervised transfer learning has been explored in detecting and classifying seven species of \textit{Eimeria} spp. in chickens ~\cite{HE2023102459} through the analysis of curated large-scale microscopic imagery~\cite{article,https://doi.org/10.48550/arxiv.2207.01419}. Data augmentation techniques, such as image flipping, adding Gaussian noise and histogram normalization, and transfer learning from ImageNet~\cite{russakovsky2015imagenet} pre-trained models have enabled the training of large deep learning models for ova detection. However, the dependency on large-scale training data was not alleviated, which limits their generalization to other biomedical applications. Weakly supervised approaches such as those based on Multiple Objects Feature Fusion (MOFF)~\cite{manescu2020weakly} and traditional image processing techniques~\cite{arco2015digital} have been used to reduce the dependency on densely annotated data, yet still assume access to large-scale datasets for learning associations between the input and target classifications. 
Advances in generative models such as GANs provided a viable mechanism to generate additional training data for \textbf{learning from limited data.} DADA~\cite{zhang2019dada} explores training with small samples from Cifar-10 and SVNH. They use GANs to augment the original dataset for more diversity with the same labels. Barz \textit{et al.}~\cite{barz2020deep} showed that the Cosine Loss is better than categorical cross-entropy whenever only a few samples are present per class. Ishikawa \textit{et al.}~\cite{ishikawa2022boosting} use conditional GAN for augmentation to improve efficiency in generating training samples but increase the computational cost. Brigato \textit{et al.}~\cite{dravid2022investigating} use Auxialiary-Classifier GANs for image synthesis and classification in low data settings. Meta-learning approaches such as MAML~\cite{finn2017model} and Prototypical Networks~\cite{https://doi.org/10.48550/arxiv.1703.05175,li2020prototypical} have enabled \textit{few-shot} learning where only a few samples per class are required for making inferences. However, such or similar approaches ~\cite{li2020prototypical,sung2018learning,chuang2020debiased} assume a reasonably large training corpus exists to create ``meta-tasks'' for learning robust representations for downstream classification. 

\section{ProtoKD: Learning from Scarce Data}\label{sec:proposed}
\textbf{Problem Statement.} In this work, we consider the task of learning from extremely scarce samples $n \leq 5$ per class. During training, the model can access a set of $N$ training examples $X{=} \{x_1, x_2, x_3, \ldots x_N\}$ drawn from $C$ classes with $n$ samples each. 
The model is presented with samples from any of the $C$ classes at test time. In contrast, meta-learning approaches~\cite{https://doi.org/10.48550/arxiv.1703.05175,NIPS2016_90e13578,finn2017model} consider the training and evaluation phases to consist of $c$-way classification tasks, where $c\ll C$. Our setup is more challenging since we have a $C$-way classification task and have access to extremely scarce data. 
\begin{figure*}[t]
\centering

\begin{tabular}{ccccccc}

\hline
\includegraphics[width=0.11\linewidth]{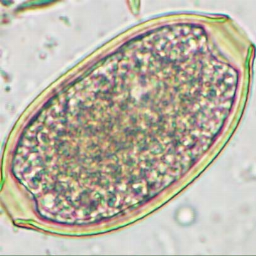} & \includegraphics[width=0.11\linewidth]{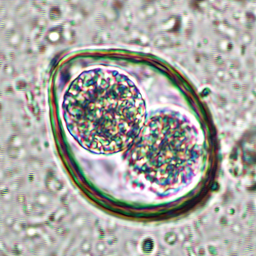} &
\includegraphics[width=0.11\linewidth]{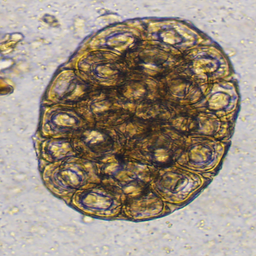} & \includegraphics[width=0.11\linewidth]{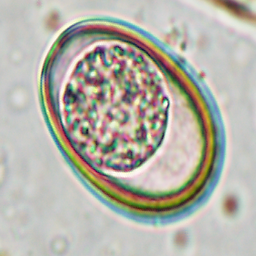} & \includegraphics[width=0.11\linewidth]{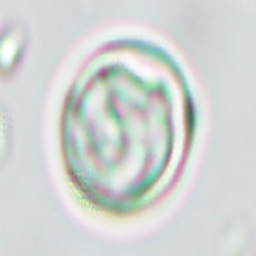} & \includegraphics[width=0.11\linewidth]{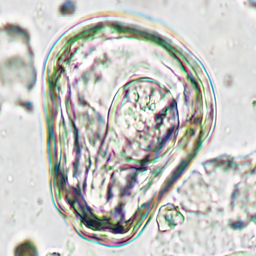} & \includegraphics[width=0.11\linewidth]{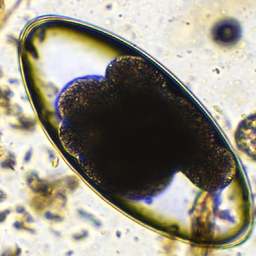} \\
\hline
Capillarids & \textit{Cystoiso-}  & \textit{Dipylidium} & \textit{Eimeria}  & \textit{Giardia}  & \textit{Moniezia} & \textit{Nematodirus} \\

(190) & \textit{spora} sp.(76) & \textit{caninum} (26) & sp. (65) &  sp. (149) & sp.(134) & sp.(53)\\
\hline
\includegraphics[width=0.11\linewidth]{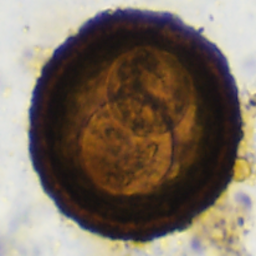} &
\includegraphics[width=0.11\linewidth]{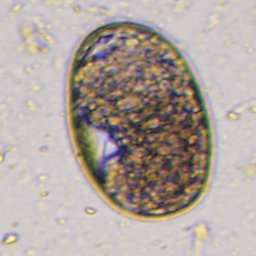} & \includegraphics[width=0.11\linewidth]{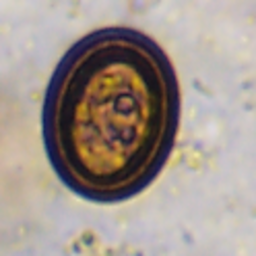} & \includegraphics[width=0.11\linewidth]{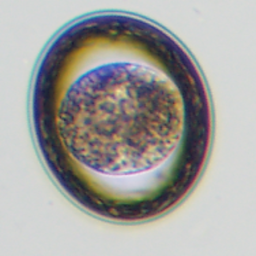} & 
\includegraphics[width=0.11\linewidth]{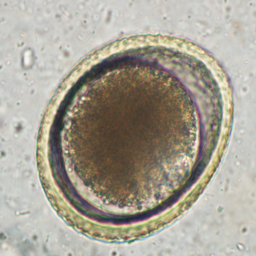} &\includegraphics[width=0.11\linewidth]{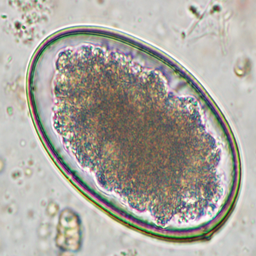} & \includegraphics[width=0.11\linewidth]{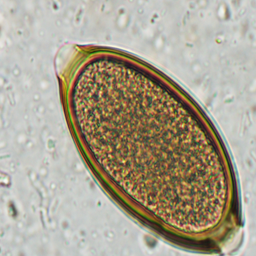}  \\
\hline
\textit{Parascaris} & Strongyles & \textit{Taeniid} & \textit{Toxascaris} & \textit{Toxocara}  & Trichostron- & \textit{Trichuris} \\

sp. (173) & (84) & eggs (90) & \textit{leonina} (32) & sp. (201) & gyles (264) & sp. (36) \\
\hline
\end{tabular}

\caption{\textbf{Dataset Characteristics.} Our parasite ova dataset has $1573$ instances across $14$ ova classes, imaged from $594$ clinical samples. An exemplar image from each class and the number of samples in each class are presented. It is highly imbalanced, with high inter- and intra-class variation. 
}
\label{fig:qual}
\end{figure*}

\subsubsection{Prototypical Networks for Scarce Data}\label{sec:proto}
Prototypical networks aim to construct a $D$-dimensional vector representation of each class, called the \textit{prototype}, that captures its underlying characteristics. Ideally, each prototype $p_j$ represents the typical example from that class and captures the intra-class variations through an embedding function $f_\theta: x_i\longrightarrow F_i$ that projects a sample $x_i$ to a $D$-dimensional vector $F_i$. The prototype is computed as the mean representation of all $n$ training samples in the class $c_j\in C$. In a typical meta-learning setup, a \textit{support} set $S$ consisting of $k$ examples from $c$ classes is sampled to construct these prototypes. A function $\phi$ provides the distance between each example in the \textit{query} set $Q$ and each prototype. A softmax function over these distances provides a probability distribution to label each example $x_i \in Q$. This setup assumes that (i) each sample in the query comes from the $c$ classes sampled in the episode and (ii) there exists a reasonably large dataset to sample the support and query sets during training to learn good representations for $c$-way classification. However, in biomedical applications, the number of training samples can be sparse due to the high data acquisition cost or the need to adapt rapidly to a changing scenario. Similarly, assuming a smaller subset of classes at inference is unrealistic since this requires prior knowledge about each test sample. 

We extend the prototypical network formulation to the extremely scarce data regime to overcome these limitations by proposing to learn the embedding function $f_\theta$ through domain-specific data augmentation. Given a support set $S_i$, each example $x_i$ is augmented through a controlled, pre-determined scheme to create $k$ samples $\{x^\prime_1, x^\prime_2, x^\prime_3, \ldots x^\prime_k\}$ that become part of the query set. Hence, each minibatch consists of examples from \textit{all} $C$ classes, with the prototypes constructed from original, uncorrupted samples and the query samples providing samples that cover the plausible intra-class variations for each class. The training objective is to minimize the negative log probability of each query sample $x^\prime_i$ belonging to its true class $c_j$ represented by its prototype $p_j$. Hence, we minimize the matching loss ($\mathcal{L}_m$) given by 
\begin{equation}
    -log p(y=j\vert x^\prime_i, \theta) = -log\left(\frac{-exp(\phi(f_\theta (x^\prime_i, p_j)))}{\sum^C_k {exp(\phi(f_\theta (x^\prime_i, p_k))}}\right)
    \label{eqn:matchLoss}
\end{equation}
where $x^\prime_i$ is an example from the query set $Q$; $\phi(\cdot)$ is a distance function that provides the distance (in the range $[0,+\infty)$) between each sample $x^\prime_i$ and a prototype $p_j$; and $f_\theta$ is an embedding function, defined as a Wide ResNet~\cite{zagoruyko2016wide}, and $\phi(\cdot)$ is the Euclidean distance. The data augmentation scheme is specific to each use case and must be designed based on prior, domain-specific knowledge. In our experiments with the parasite ova data, we apply augmentation mechanisms such as random zoom, rotation, contrast change,  flip, shear, and solarization. However, these are not plausible augmentation schemes for the genome data to capture the intra-class variations. Hence, we design genome-specific augmentation schemes based on base flipping to simulate observation error~\cite{laver2015assessing}. We add noise drawn from a normal distribution (with $0$ mean and variance of $1$) to $5\%$ of the pixels randomly selected symmetrically along the diagonal. This augmentation mechanism mimics the observation errors commonly found in genome sequencing~\cite{laver2015assessing,indla2021sim2real} and provides a natural augmentation scheme to capture the intra-class variation and additionally helps preserve the symmetry of the pseudo-image-based k-mer representations. 
We refer the reader to ~\cite{aakur2021mg} for more details on the pseudo-imaging for genomics. 

\subsubsection{Learning Intra-class Variations with Self-Distillation}\label{sec:selfdistill}
The second step is to enhance the prototypes $P{=}\{p_1, p_2, p_3, \ldots p_C\}$ to capture the intra-class variations. First, we generate \textit{pseudo-labels} for each query sample $x^\prime_i \in Q$ using the probability distribution defined in Equation~\ref{eqn:matchLoss} to create labels for the augmented samples. This \textit{self-labeled} data is then used to fine-tune the encoder network using a distillation loss given by
\begin{equation}
    \mathcal{L}_s = D_{KL}(p_t(y=j\vert x_i, \theta_k)/\tau, p_s(y=j\vert x_i, \theta_s))
    \label{eqn:distill_loss}
\end{equation}
where $D_{KL}(\cdot)$ is the Kullback–Leibler divergence~\cite{joyce2011kullback} between the probability distributions of the prototypical network defined in Section~\ref{sec:proto} and a linear layer trained on top of representations $F_i$ from $f_\theta$; $\tau$ is a temperature parameter that controls how closely the linear layer's output should match that of the prototypical network; and $\theta_t$ and $\theta_s$ are the trainable parameters of the prototypical network (i.e., the embedding function $f_\theta$) and the linear layer, respectively. We set $\tau$ to 5, chosen based on a grid search between $0$ and $10$. 
This formulation could be extended to leverage unlabeled data into a semi-supervised learning setting in future works. 
We leave that exploration to future work and focus only on the extremely scarce data setting. 

In addition to the distillation loss described above, we introduce a simple discriminative loss to help increase the separability of the decision boundary between the classes. To this end, we employ a similarity-based contrastive loss that reduces the difference between the features of each query sample $x^\prime_i$ and its corresponding prototype $p_j$ while increasing the distance to other prototypes. The intuition behind this loss is that increasing the distance between the query sample and other prototypes can capture the variability within each class since we widen its decision boundary based on its prototype $p_j$. We define this to be a discriminative loss given by
\begin{equation}
    \mathcal{L}_d = -log \left( 
    \frac{exp(p_j^T \cdot F_i)}
    {\sum^C_{k=1} {1\!\rm l}_{j \neq k} exp(p_k^T \cdot F_i)}
    \right)
    \label{eqn:disc_loss}
\end{equation}
where $F_i$ is the feature representation of a query sample $x^\prime_i$; ${1\!\rm l}\in \{0, 1\}$ indicates whether the sample is from its true class $c_j$; and both $p_j$ and $F_i$ are $\ell_2$ normalized vectors. This formulation allows us to leverage our proposed domain-specific augmentation setup for a contrastive learning mechanism without having to sample new positive and anchor examples or triplet mining, as with other contrastive learning mechanisms such as SimCLR~\cite{chen2020simple} or triplet losses~\cite{ge2018deep}.

\begin{table}[t]
\caption{\textbf{Quantitative results} on the parasite ova dataset. One example from each class was randomly sampled for training. Average performance from $10$ runs is reported.}

\centering
\resizebox{\columnwidth}{!}{
\begin{tabular}{|l|c|c|c|c|c|c|}
\hline
 & \multicolumn{2}{c|}{\textbf{Supervised}} & \multicolumn{2}{c|}{\textbf{ProtoNet}} & \multicolumn{2}{c|}{\textbf{ProtoKD}} \\ \cline{2-7} 
 & \textbf{Precision} & \textbf{Recall} & \textbf{Precision} & \textbf{Recall} & \textbf{Precision} & \textbf{Recall} \\ \hline
\textbf{Average} & 0.436 & 0.448 & 0.472 & {0.494} & \textbf{0.523} & \textbf{0.544} \\ \hline
\hline
\textbf{Capillarids} & 0.558 & 0.468 & 0.554 & \textbf{0.680} & \textbf{0.730} & 0.460 \\ \hline
\textbf{\textit{Cystoisospora} spp.} & 0.166 & 0.140 & 0.174 & 0.188 & \textbf{0.360} & \textbf{0.260} \\ \hline
\textbf{\textit{Dipylidium caninum}} & 0.116 & 0.424 & \textbf{0.236} & \textbf{0.448} & 0.080 & 0.180 \\ \hline
\textbf{\textit{Eimeria} spp.} & 0.292 & 0.152 & 0.120 & 0.222 & \textbf{0.480} & \textbf{0.290} \\ \hline
\textbf{\textit{Giardia} spp.} & 0.908 & 0.978 & 0.934 & 0.918 & \textbf{0.945} & \textbf{1.000} \\ \hline
\textbf{\textit{Moniezia} spp.} & \textbf{0.528} & \textbf{0.580} & 0.492 & 0.528 & 0.515 & 0.575 \\ \hline
\textbf{\textit{Nematodirus} spp.} & 0.272 & 0.414 & \textbf{0.352} & 0.464 & 0.250 & \textbf{0.790} \\ \hline
\textbf{\textit{Parascaris} spp. }& 0.768 & 0.746 & \textbf{0.858} & 0.716 & 0.755 & \textbf{0.785} \\ \hline
\textbf{Strongyles} & 0.268 & 0.464 & 0.362 & 0.484 & \textbf{0.565} & \textbf{0.650} \\ \hline
\textbf{\textit{Taeniid} spp.} & 0.574 & 0.334 & 0.650 & 0.320 & \textbf{0.785} & \textbf{0.715} \\ \hline
\textbf{\textit{Toxascaris leonina}} & 0.226 & 0.670 & \textbf{0.342} & \textbf{0.834} & 0.255 & 0.630 \\ \hline
\textbf{\textit{Toxocara} spp.} & 0.738 & 0.438 & 0.790 & 0.592 & \textbf{0.795} & \textbf{0.610} \\ \hline
\textbf{Trichostrongyles} & 0.562 & 0.174 & 0.560 & 0.180 & \textbf{0.665} & \textbf{0.265} \\ \hline
\textbf{\textit{Trichuris} spp.} & 0.150 & 0.300 & \textbf{0.184} & 0.322 & 0.140 & \textbf{0.400} \\ \hline
\end{tabular}
}
\label{tab:ova_results}
\end{table}

\textbf{Implementation Details.} The framework is trained end-to-end in two alternating phases. Every epoch consists of $10$ iterations of training only with the matching loss, followed by $10$ iterations of training using both distillation and discriminative losses. The matching loss cycle allows us to build prototypes of each class, while the self-distillation phase allows us to refine the prototypes by capturing the intra-class variation explicitly.  
We use a Wide-ResNet~\cite{zagoruyko2016wide}, trained from scratch, as the backbone for our mapping function, with a depth of $28$ layers, a width of $2$, and a dropout rate of $0.3$. The features are projected to a $128$-dimension vector using a linear layer. The images are resized to $128\times 128$ and pre-processed as done in ResNet~\cite{he2016deep}. 
All networks are trained for 100 epochs with a support set size of $1$ and a query set of $5$ and converge in 90 minutes.
All experiments were conducted on a workstation server with an AMD ThreadRipper CPU with 64 cores, 128 GB RAM, and an NVIDIA Titan RTX (24GB). 

\section{Experimental Evaluation}\label{sec:results}
In this section, we present the experimental setup and evaluation results for the proposed ProtoKD approach. We begin by describing the data collection process to curate the parasitic ova recognition dataset, one of the first attempts at a comprehensive benchmark for the task. We discuss the metrics and baselines used for evaluation and present the quantitative results. We conclude by demonstrating the generalization capabilities of the proposed ProtoKD framework to other biomedical applications, such as genome classification.

\begin{table}[t]
    \centering
    \caption{\textbf{Ablation studies} to determine the impact of each component on the performance of ProtoKD on the parasite ova recognition task.}
    \begin{tabular}{|c|c|c|c|c|c|}
    \toprule
    \textbf{$\mathcal{L}_m$} & \textbf{$\mathcal{L}_s$} & \textbf{$\mathcal{L}_d$} & \textbf{Precision} & \textbf{Recall} & \textbf{F1-Score}\\\hline
     \ding{51} &  \ding{55} &  \ding{55} & 0.472 & 0.494 & 0.483 \\\hline
     \ding{55} &  \ding{51} &  \ding{55} & 0.436 & 0.448 & 0.442 \\\hline
     \ding{51} &  \ding{51} &  \ding{55} & 0.497 & 0.501 & 0.499 \\\hline
     \ding{51} &  \ding{55} &  \ding{51} & 0.508 & 0.516 &  0.512 \\\hline
     \ding{51} &  \ding{51} &  \ding{51} & \textbf{0.523} & \textbf{0.544} &  \textbf{0.533} \\\hline
    \end{tabular}
    \label{tab:ablation}
\end{table}

\textbf{Data Collection.} We collected 1573 ova examples, curated from 594 clinical samples at a local [redacted for anonymity] diagnostics laboratory. 
Each type of egg was collected by performing a centrifugal fecal flotation on samples from various hosts infected with different parasites and were subsequently observed under a microscope by a certified parasitologist. Following identification, images were captured using the Olympus BX43 microscope (Tokyo, Japan) and the Olympus Cell Sens Entry software v1.18. 
Based on their frequency of occurrence in samples received at both local and national diagnostics labs, the following 14 parasitic ova were considered: Capillarids, \textit{Cystoisospora} spp., \textit{Dipylidium caninum}, \textit{Eimeria} spp., \textit{Giardia} spp., \textit{Moniezia} spp., \textit{Nematodirus} spp., \textit{Parascaris} spp., Strongyles, \textit{Taeniid eggs}, \textit{Toxascaris leonina}, \textit{Toxocara} spp., Trichostrongyles, and \textit{Trichuris} spp. Figure~\ref{fig:qual} provides examples and statistics. 
Each parasite and its relative size determined where the magnification would be 40x, 20x, or 10x objective. 
Most of the images captured are at 10x objective magnification which is available on a large majority of microscopes. Clinical samples were collected until each class had at least 25 examples to provide a comprehensive benchmark for evaluating machine learning frameworks for ova recognition with extremely scarce data. 

\textbf{Metrics and Baselines.} Due to the highly imbalanced nature of the data, we choose precision and recall per class as our evaluation metric since accuracy can be highly skewed towards classes with more examples. We evaluate the performance of each algorithm with a training set with $1$ example per class and report the mean results from $10$ random trials to avoid conflating the results due to the choice of the example from each class. 
We choose a fully supervised Wide-ResNet as our backbone network for all baselines, including a fully supervised one, the modified ProtoNet from Section~\ref{sec:proto}, and ProtoKD. 
All hyperparameters for each baseline are kept constant for each trial and trained for $100$ epochs with early stopping based on a validation set of $5$ samples per class.

\textbf{Performance on Parasite Ova Data.} We first evaluate and compare our approach against baselines on the parasite ova dataset. Table~\ref{tab:ova_results} summarizes the results when training with only one example per class. The proposed ProtoKD approach performs well across classes, with an average precision of $0.523$ and recall of $0.544$, outperforming the supervised and ProtoNet baselines. Of particular interest is the performance of the baselines on the two classes with the least intra-class variation (\textit{Giardia} spp.) and the highest intra-class variation (\textit{Nematodirus} spp.). With \textit{Nematodirus} spp., ProtoKD's recall ($0.790$) was almost $1.5$ times that of ProtoNet ($0.464$) and the supervised ($0.414$) baselines, which indicates that the self-distillation and discriminative losses played their role in learning robust prototypes. 
All three baselines performed well on Giardia, with ProtoKD achieving $100\%$ recall with a high precision of $0.945$. 
On average, we find that the ProtoKD achieves high recall at the cost of precision, particularly in the case of classes with a large number of samples, such as \textit{Parascaris} spp. and \textit{Trichuris} spp. We hypothesize that this is an effect of the contrastive, discriminative loss intended to expand each class's decision boundary. Interestingly, the ProtoKD severely fails on \textit{Dipylidium caninum}. Upon close inspection, \textit{Dipylidium caninum} was highly confused with Tirchostrongyle eggs, which, though visually similar  (see Figure~\ref{fig:qual}), are functionally different.
%
The strong augmentation scheme resulted in overlapping decision boundaries due to the extreme variation in Trichostrongyle examples. We anticipate using super-resolution mechanisms~\cite{liao2022comparative,zhu2023attention} to enhance the images will help find better representations. 

\begin{table}[t]
\caption{\textbf{Evaluation on metagenome classification.} Host and average pathogen f1-scores are used following prior work~\cite{aakur2021mg}. 
    }
    \centering
    \resizebox{\columnwidth}{!}{
    \begin{tabular}{|c|c|c|c|c|c|c|}
        \hline
        \multirow{2}{*}{\# Samples} & \multicolumn{2}{c|}{\textbf{MG-NET}} & \multicolumn{2}{c|} {\textbf{ProtoNet}} & \multicolumn{2}{c|}{\textbf{ProtoKD}}\\
        \cline{2-7}
        & Host F1 & Pathogen F1 & Host F1 & Pathogen F1 & Host F1 & Pathogen F1\\
        \hline
        1 & 0.230 &  0.026 & 0.596 & 0.108 & \textbf{0.699}  & \textbf{0.121}\\
        5 & 0.330 & 0.029 & 0.690 & 0.127 & \textbf{0.804} & \textbf{0.139} \\
        10 & 0.320 & 0.035 & \textbf{0.840} & 0.081 & 0.780 & \textbf{0.129} \\
        15 & 0.360 & 0.047 & \textbf{0.850} & 0.109 & 0.824 & \textbf{0.127} \\
        20 &  0.370 & 0.051 & 0.780 & 0.147 & \textbf{0.799} & \textbf{0.151} \\
        25 & 0.371 & 0.056 & 0.720 & \textbf{0.200} &  \textbf{0.775} & 0.193 \\
        \hline
    \end{tabular}
    }
    
    \label{tab:genome_results}
\end{table}

\textbf{Ablation Studies.} 
We perform ablation studies to systematically evaluate the impact of the different components of the approach on its performance. The three loss functions, defined in Equations 1, 2, and 3, are the major components of the approach. Hence, we ablate over the impact of using $\mathcal{L}_s$ and $\mathcal{L}_d$ in combination with $\mathcal{L}_m$ and report results in Table~\ref{tab:ablation}. Using only the matching loss (Equation 1), the approach degenerates to a standard ProtoNet, one of our baselines (Row 1). When $\tau$ in $\mathcal{L}_s$ (defined in Equation 2) is set to 1, it becomes the standard cross-entropy loss and hence is equivalent to our fully supervised baseline (Row 2). It can be seen that using knowledge distillation loss ($\mathcal{L}_c$) alone or matching loss ($\mathcal{L}_m$) performs reasonably well, although not as much as the proposed ProtoKD. Adding the discriminative loss ($\mathcal{L}_d$) with the matching loss ($\mathcal{L}_m$) provides a higher increase in performance than combining the matching loss ($\mathcal{L}_m$) with self-distillation ($\mathcal{L}_s$). Combining all three provides a higher increase overall, indicating the subtle balance between learning inter-class and intra-class variations provided by the alternating training methodology proposed in ProtoKD. Note that this formulation can naturally be extended to semi-supervised learning where the self-distillation loss ($\mathcal{L}_m$) can be used to train on unlabeled data. We leave that to future work since our focus is on tackling the problem of learning from scarcely available ($<5$ samples per class) training data. 

\textbf{Extension to Other Biomedical Applications.} In addition to our experiments on parasite ova recognition, we evaluate the generalizability of the proposed ProtoKD formalism to other biomedical applications by evaluating its ability to learn representations from an entirely separate application: metagenome sequences. We evaluate the ProtoKD framework on the data provided by MG-NET, which has $31,580$ sequence reads across seven classes - Bovine (host), \textit{B. trehalosi}, \textit{H. somni}, \textit{M. bovis}, \textit{M. haemolytica}, \textit{P. multocida} and \textit{T. pyogenes}. Specifically, we use the pseudo-images generated by the MG-NET framework as input and evaluate it by training with varying samples per class. The test set was fixed with 8192 samples for a fair comparison with MG-NET in all the scenarios. Average performance from 10 trials is reported. 
Table~\ref{tab:genome_results} summarizes the result. We can see that the proposed ProtoKD framework and ProtoNet outperform the supervised MG-NET at very low samples, i.e., less than 25 samples per class. It takes MG-NET at least 500 samples per class to outperform ProtoKD with 25 samples, achieving an overall host F1-score of $0.906$ and an average pathogen F1-score of $0.319$. 
Interestingly, the performance initially reduces as the number of samples per class is increased. We attribute this phenomenon to the fact that fine-grained recognition requires \textit{highly distinct} samples for learning robust features. Genomes between closely related species have shown to have similar genome sequences~\cite{aakur2021metagenome2vec}; hence, larger amounts of data do not necessarily translate into better performance. 
For example, ProtoNet achieves a higher pathogen F1-score ($0.200$) than ProtoKD ($0.193$) at 25 samples. It has $100\%$ precision and $0\%$ recall on two pathogen classes, indicating that the model does not make balanced predictions and fails on edge cases. We anticipate that including structural information~\cite{aakur2021mg,aakur2021metagenome2vec} and other metadata will improve the performance. 

\section{Conclusion and Future Work}
In this work, we presented \textit{ProtoKD}, one of the first works to tackle the problem of learning from extremely scarce training samples. Using a benchmark dataset of parasitic ova, we demonstrate its strong ability to learn robust representations from just one example per class. Experiments on large-scale metagenome-based taxonomic profiling data demonstrated its generalizability to other downstream applications. We anticipate using super-resolution to enhance the images will help find better representations. We aim to extend this framework for scaling deep learning frameworks to work with highly constrained data typical in biomedical applications such as disease diagnostics and the Internet of Medical Things. 

\textbf{Acknowledgement.} This work was partially supported by the US National Science Foundation (NSF) grant IIS 1955230. 

\bibliographystyle{IEEEtran}
\bibliography{egbib}
\end{document}